\documentclass[9pt,twocolumn,twoside]{osajnl}
\usepackage{graphicx}
\usepackage{amsmath}
\usepackage{algorithm}
\usepackage{algpseudocode}
\usepackage{pifont}
\usepackage{epsfig,amsfonts}
\usepackage{amssymb}
\usepackage{amsthm}
\usepackage{multirow}
\usepackage{tabularx}
\usepackage{xcolor}
\usepackage{array}
\usepackage{url}
\usepackage[justification=justified,format=plain]{caption}

\journal{optica} 

\DeclareMathOperator*{\argmin}{arg\,min}
\DeclareMathOperator*{\argmax}{arg\,max}
\DeclareMathOperator*{\mean}{mean}

\setboolean{shortarticle}{false}

\title{Power-Balanced Hybrid Optics Boosted Design for Achromatic Extended-Depth-of-Field Imaging via Optimized Mixed OTF}

\author[1,*]{Seyyed Reza Miri Rostami}
\author[1]{Samuel Pinilla}
\author[1]{Igor Shevkunov}
\author[1]{Vladimir Katkovnik}
\author[1]{Karen Egiazarian}

\affil[1]{Computing Sciences Unit, Faculty of Information Technology and Communication Sciences, Tampere University, FI-33720 Tampere, Finland.}

\affil[*]{Corresponding author: SeyyedReza.MiriRostami@tuni.fi}

\begin{abstract}
The power-balanced hybrid optical imaging system is a special design of a diffractive computational camera, introduced in this paper, with image formation by a refractive lens and Multilevel Phase Mask (MPM). This system provides a long focal depth with low chromatic aberrations thanks to MPM and a high energy light concentration due to the refractive lens. We introduce the concept of optical power balance between the lens and MPM which controls the contribution of each element to modulate the incoming light. Additional unique features of our MPM design are the inclusion of quantization of the MPM's shape on the number of levels and the Fresnel order (thickness) using a smoothing function. To optimize optical power-balance as well as the MPM, we build a fully-differentiable image formation model for joint optimization of optical and imaging parameters for the proposed camera using Neural Network techniques. Additionally, we optimize a single Wiener-like optical transfer function (OTF) invariant to depth to reconstruct a sharp image. We numerically and experimentally compare the designed system with its counterparts, lensless and just-lens optical systems, for the visible wavelength interval (400-700)~nm and the depth-of-ﬁeld range (0.5-$\infty$~m for numerical and 0.5-2~m for experimental). 
The attained results demonstrate that the proposed system equipped with the optimal OTF overcomes its counterparts (even when they are used with optimized OTF) in terms of reconstruction quality for off-focus distances. The simulation results also reveal that optimizing the optical power-balance, Fresnel order, and the number of levels parameters are essential for system performance attaining an improvement of up to 5dB of PSNR using the optimized OTF compared with its counterpart lensless setup.
\end{abstract}

\setboolean{displaycopyright}{true}
\begin{document}
\maketitle
\section{Introduction}\label{sec:introduction}\vspace{-0.5em}
Computational imaging with encoding flat diffractive optical elements (DOEs) (e.g., binary/multi-level phase elements \cite{leveque2020co,baek2020end}; meta-optical elements included \cite{chen2020flat}) is a multidisciplinary research field in the intersection of optics, mathematics, and digital image processing. This particular exciting academic intersection has an increasing interest in designing DOEs for applications such as computational photography \cite{sitzmann2018end,dun2020learned}, augmented reality \cite{krajancich2020factored}, spectral imaging \cite{10.1145/3306346.3322946}, microscopy \cite{adams2017single}, among others that are leading the need for highly miniaturized optical systems \cite{antipa2018diffusercam,yanny2019miniature}. The design of imaging setups for the aforementioned applications involves the optimization of optical elements such as amplitude masks \cite{pinilla2018coded,correa2016spatiotemporal,bacca2019super,jerez2020fast}, refractive lenses \cite{kuo2020chip}, DOEs \cite{baek2020end}, diffusers \cite{antipa2018diffusercam}, various types of phase masks/plates ~\cite{boominathan2020phlatcam}, etc. In this work, the elements of interest to be jointly designed are refractive lens and DOE in order to improve the depth-of-field (DoF) and reduce the chromatic aberrations of the system, problem that is also known as achromatic extended-depth-of-field (EDoF) imaging.

Two basic approaches are exploited to effectively design and develop all-in-focus optical imaging. The first one deals with the design of optical systems which are highly insensitive to a distance between the object and camera. Contrary to it, another approach aims at the design of optical systems which are highly sensitive to variations of this distance. The depth map of the scene is reconstructed and used for all-in-focus imaging \cite{haim2015computational}, or to control physical parameters of the system using tunable and programmable optical devices\cite{Li:18,9064909,cossairt2010spectral}. In this work, we follow the mainstream of the first approach to develop photography cameras and the corresponding reconstruction algorithm for sharp and high-quality imaging with achromatic EDoF.

Here we propose an optical power-balanced hybrid setup which is a composition of a refractive lens and a multilevel phase mask (MPM) as DOE to extend the DoF capabilities of a camera. We include the number of levels to control the quantization level and the Fresnel order which tune the thickness (Manufacturing aspects) in the MPM design, unique features that are flexible to be optimized. This system introduces the concept of optical power-balance between the lens and MPM, which controls the contribution of each element to modulate the incoming light. We mathematically motivate the use of our optical power-balance hybrid system by providing analytical remarks showing that they allow better estimation of an image in terms of accuracy, visual perception, and less chromatic aberration over a wide DoF compared to setups without power-balanced optics that employ either lens or lensless (just DOE). 

To optimize the optical power-balance and MPM, we build a fully differentiable image formation model for joint optimization of optical and imaging parameters for the designed computational camera using neural networks. In particular, for the number of levels and Fresnel order features we introduce a smoothing function because both parameters are modeled as piecewise continuous operations. Our optimization framework, in contrast to the previous works which have pursued to achieve narrow point-spread functions along the depth to facilitate the inverse imaging, designs a single Wiener-like optical transfer function (OTF) to estimate a scene from blurred observations invariant to depth. Moreover, BM3DSHARP algorithm \cite{Dabov07jointimage} is included for sparsity modeling as a prior for images to be reconstructed. Numerical results reveal that the optimization of the optical power balance, Fresnel order, and the number of levels parameters are essential for system performance attaining an improvement up to 5dB of PSNR using the optimized OTF compared with its counterpart lensless setup. Additionally, we point out that due to the new included variables in the MPM design (number of levels, thickness, and power-balance), the number of possible combinations to compare the proposed system with the state-of-the-art is huge and in order to avoid building several MPMs to physically analyze the performance of our camera, we build a setup which exploits the phase capabilities of a spatial light modulator (SLM) to investigate the performance of the proposed MPM design.

The contribution of this work can be summarized as follows.
\begin{itemize}
	\item The power-balanced hybrid optical system with optimized power-balance between the refractive and diffractive elements is presented and studied;\vspace{-0.5em}
	\item Inclusion of Fresnel order and the number of levels in the optimization pipeline of MPM via smoothing function;\vspace{-0.5em}
	\item Novel single Wiener-like OTF is proposed for inverse imaging from blurred observations invariant to depth;\vspace{-0.5em}
	\item The optimal parameters of the system and MPM are produced under end-to-end framework solving multi-objective optimization problem with PSNR as criterion function;\vspace{-0.5em}
	\item The performance of the proposed optical setup is demonstrated by numerical simulation and experimental tests with SLM for implementation of MPM.
\end{itemize}

	\begin{table*}[t!]
	\centering
	\caption{\small Comparison of different phase modulation setups for EDoF enabling the state-of-the-art. We show the design methods and optimizing physical variables. The mathematical models for the phase distribution of the optics are included. Also, manufacturing factors essential for the lens/DOE elements are mentioned, if any exist.}
	\footnotesize
	\begin{tabular}{m{0.05cm} m{2cm} m{1.8cm} m{3.2cm} m{3.5cm} m{3cm} m{2cm}}
		\hline
		\hline 
		\vspace{0.1em} & \textbf{Ref.} \vspace{0.4em} & \textbf{Optics Type} \vspace{0.4em} & \textbf{Mathematical Model} \vspace{0.4em}& \textbf{Phase Model} \vspace{0.4em}& \textbf{Manufacturing factors} \vspace{0.4em} & \textbf{Design Method} \vspace{0.4em}\\
		\hline 
		1 & 1995:\cite{dowski1995extended} & Lens \& Phase Mask & $e^{-i\frac{\pi}{\lambda f} (x^{2} + y^{2})} \times e^{i\beta (x^{3} + y^{3})}$ & \multicolumn{1}{c}{Cubic} & \multicolumn{1}{c}{-} & Analytical \\
		\hline
		2 & 2004:\cite{flores2004achromatic,sherif2004phase}
		
		2007:\cite{yang2007optimized}
		
		2009:\cite{zhou2009rational}
		
		2020:\cite{9362315} & Lens \& Absolute Phase & $e^{-i\frac{\pi}{\lambda f} (x^{2} + y^{2})}\times e^{i\varphi(x,y)}$ & Parametric (Bessel polynomials, logarithmic or rings are employed) & \multicolumn{1}{c}{-}& PSF Engineering.;
		OTF Engineering.; 
		End-to-End \\
		\hline 
		3 & 2008:\cite{ben2008optimal}
		
		2010:\cite{milgrom2010pupil}
		
		2015:\cite{haim2015computational,burcklen2015experimental}
		
		2017:\cite{ryu2017design}
		
		2018:\cite{elmalem2018learned}
		
		2019:\cite{fontbonne2019experimental}
		& Lens \& Binary Phase Mask & $e^{-i\frac{\pi}{\lambda f} (x^{2} + y^{2})}\times e^{i\mathcal{B}(\varphi(x,y))}$ & Binary (Rings are employed) & \multicolumn{1}{c}{two number of levels} & PSF Engineering; End-to-End \\
		\hline
		4 & 2007:\cite{garcia2007design}
		
		2010:\cite{cossairt2010diffusion}& Diffuser &$e^{i\varphi(x,y)}$ & Parametric (Rings are employed) & \multicolumn{1}{c}{-} & PSF Engineering; Analytical \\
		\hline
		5 & 2018:\cite{sitzmann2018end}
		
		2020:\cite{jin2020deep} & Phase Mask & $ e^{i\sum_{r=1}^{R}\rho_{r}P_{r}(x,y)}$ & Free-shape (Zernike/Fourier Basis, $P_{r}(x,y)$, are employed) & \multicolumn{1}{c}{-} & End-to-End \\
		\hline
		6 & 2008:\cite{caron2008polynomial}
		
		2009:\cite{zhou2009optimized}
		
		2020:\cite{banerji2020extreme,gonzalez2020jacobi}
		
		2021:\cite{9362315}
		& Phase Mask & $e^{i\varphi(x,y)}$ & Parametric (Jacobi–Fourier phase mask, rings, Bessel basis, or fringes are employed) & \multicolumn{1}{c}{-} & PSF Engineering; End-to-End;\\
		\hline 
		7 &
		
		2020:\cite{leveque2020co}
		& Binary Phase Mask & $e^{i\mathcal{B}(\varphi(x,y))}$ & Parametric (Rings are employed) & \multicolumn{1}{c}{two number of levels} & PSF Engineering \\
		\hline
		8 & 2012:\cite{sheppard2012three}
		
		2019:\cite{katkovnik2019lensless,ponomarenko2019phase}
		
		2020:\cite{baek2020end}
		
		2021:\cite{reza} 
		& Multi-level Phase Mask & $e^{i\mathcal{M}(\varphi(x,y))}$ & Parametric (Bessel basis, cubic, quadratic are employed) & \multicolumn{1}{c}{multiple number of levels} & Analytical;
		End-to-End \\
		
		\hline
		9* & \textbf{This paper}: Optical Power-Balanced Hybrid & Lens \& Multi-level Phase Mask
		& $e^{-i\frac{\pi(1-\alpha)}{\lambda f} (x^{2} + y^{2})}\times e^{i\mathcal{M}(\varphi_{\alpha}(x,y))}$ & Parametric (Quadratic, cubic and free-shape are employed) & multiple number of levels, thickness, optical power-balance & End-to-End \\
		\hline
		\hline
	\end{tabular}
	\label{tab:1}\vspace{-0.5em}
	\begin{flushright}
		$\mathcal{B}(\cdot)$/$\mathcal{M}(\cdot)$ non-linear mappings to denote binary/multi-level phase values. In Mathematical Model, the propagation phase $\frac{j\pi }{\lambda }\left( \frac{1}{d_{1}}+\frac{1}{d_{2}}\right) \left(x^{2}+y^{2}\right) $
		is omitted, see \eqref{GPF}.
	\end{flushright}\vspace{-2em}
\end{table*}

	
\textit{Scope.} We believe that the proposed framework for end-to-end optimization provides steps forward for low-level features of the optics (thickness and number of levels) and high-level contributions with the designed single Wiener-like OTF for inverse imaging. This reconstruction process, for example, could be used in photography and video applications due to the low computational cost. Additionally, the insights provided by the optical power-balance concept lead to a better understanding of how optics can be designed to improve imaging quality by proper light modulation.\vspace{-0.8em}
		
\section{Related Work}
\label{Related Work}
In Table \ref{tab:1}, we provide the relevant references to the achromatic EDoF problem of diffractive optics imaging. In this table, we present the mathematical models for phase modulation design as well as manufacturing aspects subject to optimization. We remark that our target is to build an optical system equipped with computational inverse imaging. As a result, the optics with MPM for direct focusing/imaging on the sensor are out of our consideration.
	
As primarily we focus on the MPM design, it is natural to classify them accordingly, i.e., according to the basic ideas of the approach and the design methodology. We distinguish the following three basic groups of optical models that appear in design problems: free-shape phase (often absolute phase); lensless optics; hybrid optics conceived as a composition of the lens and MPM. In the first group, the object of the design is a modulation phase often an absolute phase enabling a desirable manipulation of wavefronts. Examples are given in rows 1 and 2 of Table \ref{tab:1}.
	
In the first row, a simple model of the quadratic and cubic phases is used. The quadratic components are fixed by the parameters of the corresponding focusing lens and the variables of the design are the parameters of the cubic component. The main disadvantage of this technique is that for extreme depth values the cubic component produces strong aberrations reducing quality in the reconstruction. In the second row of the table, this cubic component is replaced by arbitrary phase functions $\varphi(x,y)$ which is a free-shape phase design.
 The alternative second group of design optical models is lensless with various phase plates/masks flat and non-flat instead of the conventional refractive lens. In Table~\ref{tab:1}, the examples of this group of optical elements can be seen in rows 4,~5,~6,~7, and 8. However, just a phase mask employed to extend the DoF of an imaging system, they suffer from strong chromatic aberrations, limiting the achievable reconstruction quality.
	
The so-called hybrids form the third group of optical models which are presented in row 3 of Table~\ref{tab:1}. In this approach, a binary phase mask using concentric rings is produced to enabling the fabrication of thin optical elements that encode the incoming light. 
However, the fact that just two levels are employed to the phase profile of the phase mask also limits the achievable DoF of the system. In the hybrids, the lens is combined with a phase mask/plate which is usually flat. Thus, the phase design is restricted by the structure parameters of mask/plate that differs these designs from those for group 1, where the free-shape phase can be arbitrary. The last row of the table is addressed to the optical setup which is the topic of this paper. It is a hybrid with DOE as special MPM with the optimized optical power balance between the lens and MPM.
	
The introduced classification of the optical elements is not one-to-one, as in particular, the prominent wavefront coding (WFC) proposed by E.R. Dowski and W.T. Cathey (1995), row 1 in the table, can be treated as a hybrid of the lens with the cubic phase mask. 
However, the existence of the lens is not so important for the methodology as the design is focused on optimization of the phase $\varphi(x,y)$, which can be arbitrary and not restricted to the cubic. The fact that the lens is used appeared as an essential point at the stage of implementation when the cubic phase can be engraved on the lens surface or presented by the cubic phase mask as an additional optical element.

A broadband imaging with phase mask is a promising technique for achromatic EDoF imaging. One of the challenges in broadband imaging with a phase mask is a strong dispersion causing significant color aberration. Nevertheless, a flow of publications demonstrates significant progress in this field of research. It is shown in the last column of Table~\ref{tab:1}, that optimization and design methods could be divided into three different frameworks: analytical, PSF engineering (fitting), and end-to-end optimization. Recently, the superiority of end-to-end optimization using convolutional neural networks for image processing and optics design is demonstrated in a number of publications \cite{elmalem2018learned,sitzmann2018end,9362315,baek2020end,khan2020flatnet,8747330,Metzler_2020_CVPR}. In terms of the introduced classification, these works are mostly belong to the first group of the phase mask models with free-shape phase design despite all differences in implementation that may concern lensless or hybrid structures. \vspace{-0.8em}
	
\section{End-to-End Optimization of Optical Power-Balanced Hybrid optics} 
\label{Power-Balanced-math}
\begin{figure}[t!]
	\centering
	\includegraphics[width=0.8\linewidth]{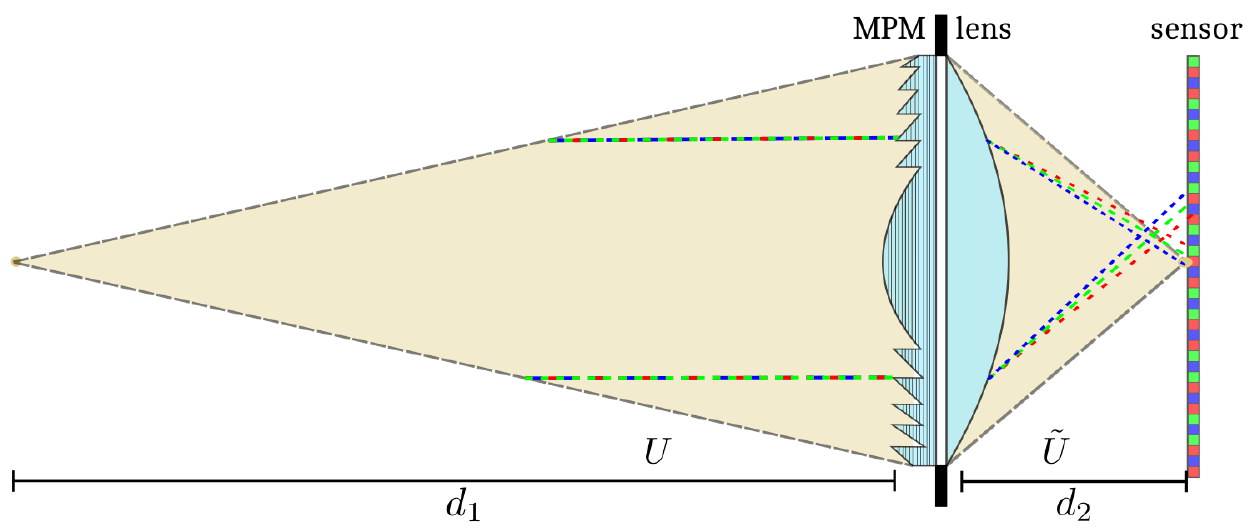}
	\caption{\small A light wave with a given wavelength and a curvature for a point source at a distance $d_{1}$ propagates on the aperture plane containing MPM (refractive index $n$) to be designed. The MPM modulates the phase of the incident wavefront. The resulting wavefront propagates through the lens to the aperture-sensor, distance $d_{2}$, via the Fresnel propagation model. The intensities of the sensor-incident wavefront define PSFs.}\vspace{-1em}
	\label{fig:system}
\end{figure}

\vspace{-0.5em}
Our proposed optical setup in Fig.~\ref{fig:system}, object, aperture, and sensor are 2D flat, where $d_{1}$ is a distance between the object and the aperture, $d_{2}$ is a distance from the aperture to the sensor ($d_{2}\ll d_{1}$), $f_{\lambda_{0}}$ is a focal distance of the optics. In what follows, we use coordinates $(\xi ,\eta )$, $(x,y)$, and $(u,v)$ for object, aperture, and sensor, respectively. \vspace{-0.6em}

\subsection{Image Formation Model}
\subsubsection{PSF-based RGB imaging}
From the Fourier wave optics theory, the response of an optical system to an input wavefront is modeled as a convolution of the system's PSF and a true object-image. Let us assume that there are both a lens and MPM in the aperture, then a generalized pupil function of the system shown in Fig.~\ref{fig:system} is of the form \cite{goodman2005introduction}
	\begin{equation}
	\mathcal{P}_{\lambda}(x,y)=\mathcal{P}_{A}(x,y)e^ {\frac{j\pi }{\lambda }\left( {\frac{1}{d_{1}}}+{\frac{1}{d_{2}}-}\frac{1}{f_{\lambda }}\right) \left( x^{2}+y^{2}\right) +j\varphi_{\lambda _{0},\lambda}(x,y)}.
	\label{GPF}
	\end{equation}		
In \eqref{GPF}, $f_{\lambda}$ is a lens focal distance for the wavelength $\lambda $, $P_{A}(x,y)$ represents the aperture of the optics and $\varphi_{\lambda _{0},\lambda }(x,y)$ models the phase delay enabled by MPM for the wavelength $\lambda $ provided that $\lambda _{0}$ is the wavelength design-parameter for MPM. In this formula, the phase $\frac{j\pi }{\lambda }\left( \frac{1}{d_{1}}+\frac{1}{d_{2}}\right) \left(x^{2}+y^{2}\right) $ appears due to propagation of the coherent wavefront from the object to the aperture (distance $d_{1}$) and from the aperture to the sensor plane (distance $d_{2}$), and $\frac{-j\pi }{\lambda f_{\lambda }} \left(x^{2}+y^{2}\right) $ is a quadratic phase delay due to the lens. For the lensless system 
	\begin{equation}
	\mathcal{P}_{\lambda}(x,y)=\mathcal{P}_{A}(x,y)e^{\frac{j\pi }{\lambda }\left( {\frac{1}{d_{1}}}+{\frac{1}{d_{2}}}\right) \left(x^{2}+y^{2}\right) +j \varphi_{\lambda _{0},\lambda }(x,y)},
	\label{GPF-lensless}
	\end{equation}
and for the lens system without MPM, $\varphi_{\lambda _{0},\lambda}(x,y)\equiv 0$ in \eqref{GPF}.

In the hybrid system, which is the topic of this paper, the optical power of the lens $1/f_{\lambda}$ is shared between the lens and the MPM and the generalized aperture takes the form 
\begin{equation}
	\mathcal{P}_{\lambda}(x,y)=\mathcal{P}_{A}(x,y)e^{\frac{j\pi }{\lambda }\left({\frac{1}{d_{1}}}+{\frac{1}{d_{2}}}-\frac{1-\alpha}{f_{\lambda }} \right) \left( x^{2}+y^{2}\right) +j\varphi_{\lambda _{0},\lambda, \alpha }(x,y)},
	\label{GPF-hybrid}
\end{equation}
where the parameter $\alpha\in [0,1]$. Observe, that $\alpha=0$ or $\alpha=1$ correspond to system without balancing light modulation between MPM and lens. In addition, the index $\alpha$ in $\varphi_{\lambda _{0},\lambda, \alpha }(x,y)$ shows that the magnitude of the quadratic component of the absolute phase used in the MPM design is defined by this parameter value supporting a proper sharing of the optical power. We point out that this parametrization of the optics phase profile is as simple as powerful because it allows changing between two different systems (hybrid and just MPM) by setting values of $\alpha$. To the best of our knowledge this is the first time that the optimization of the optics phase profile of an imaging system includes a parameter with the capability to determine the appropriate setup for the achromatic EDoF task.

The PSF of the coherent monochromatic optical system for the wavelength $\lambda$ is calculated by the formula \cite{goodman2005introduction}:
\begin{equation}
	PSF_{\lambda }^{coh}(u,v)=\mathcal{F}_{\mathcal{P}_{\lambda}}\left(\frac{u}{d_{2}\lambda },\frac{v}{d_{2}\lambda }\right),
	\label{PSF=COHERENT} 
	\end{equation}
	where $\mathcal{F}_{\mathcal{P}_{\lambda}}$ is the Fourier transform of $\mathcal{P}_{\lambda }(x,y)$. Then, PSF for the corresponding incoherent imaging, which is a topic of this paper, is a squared absolute value of $PSF_{\lambda }^{coh}(u,v)$. After normalization, this PSF function takes the form:
	\begin{equation}
	PSF_{\lambda }(u,v)= \frac{\left \lvert PSF_{\lambda }^{coh}(u,v)\right\rvert^{2}}{\iint_{-\infty }^{\infty}\left \lvert PSF_{\lambda }^{coh}(u,v)\right\rvert ^{2}dudv}. 
	\label{PSF}
\end{equation}
	
We calculate PSF for RGB color imaging assuming that the incoherent radiation is broadband and the intensity registered by an RGB sensor per $c$-band channel is an integration of the monochromatic intensity over the wavelength range $\Lambda $ with the weights $T_{c}(\lambda )$ defined by the sensor color filter array (CFA) and spectral response of the sensor. Normalizing these sensitivities on $\lambda$, i.e. $\int_{\Lambda }T_{c}(\lambda )d\lambda=1$, we obtain RGB channels PSFs
\begin{align}
	PSF_{c}(u,v) = \frac{\int_{\Lambda }PSF_{\lambda }(u,v)T_{c}(\lambda )d\lambda }{\iint_{-\infty }^{\infty}\int_{\Lambda }PSF_{\lambda }(u,v)T_{c}(\lambda )d\lambda dudv}
	\text{, } c\in \{r,g,b\}\text{,}
	\label{PSF_ave} 
\end{align}
where the monochromatic $PSF_{\lambda}$ is averaged over $\lambda$ with the weights $T_{c}(\lambda)$. 
	
Contrary to the conventional approaches for PSF-based RGB imaging, which use \eqref{PSF} with three fixed wavelengths $\lambda$ (often, 450, 550, and 650 nm) \cite{sitzmann2018end}, we take into consideration spectral properties of the sensor and in this way obtain more accurate modeling of image formation \cite{katkovnik2019lensless}. Additionally, \eqref{PSF_ave} significantly differs from the current literature because of the normalization term which we introduced. This term allows to keep the contribution of each RGB channels PSFs the same. Thus, the OTF for \eqref{PSF_ave} is calculated as the Fourier transform of $PSF_{c }(u,v)$
\begin{align}	
	OTF_{c}(f_{x},f_{y})=\iint_{-\infty }^{\infty }PSF_{c}(u,v)e^{-j2\pi (f_{x}u+f_{y}v)}dudv,
  \label{OTF1}
\end{align}
where $(f_{x},f_{y})$ are the Fourier frequency variables.

\subsubsection{From PSF to Image}
Let us introduce $PSFs$ for defocus scenarios with notation $PSF_{c,\delta }(x,y)$, where $\delta$ is a defocus distance in $d_{1}$, such that $d_{1}=d_{1}^0+\delta$ with $d_{1}^0$ equal to the focal distance between the aperture and the object. Introduce a set $\mathcal{D}$ of defocus values $\delta \in\mathcal{D}$ defining area of the desirable EDoF. It is worth noting that the corresponding optical transfer functions are used with notation $OTF_{c,\delta}(f_{x},f_{y})$. The definition of $OTF_{c,\delta}(f_{x},f_{y})$ corresponds to \eqref{OTF1}, where $PSF_{c}$ is replaced by $PSF_{c,\delta}$. Thus, let $I_{c, \delta }^s(u,v)$ and $I_{c}^o(u,v)$ be wavefront intensities at the sensor (registered focused/misfocused images) and the intensity of the object (true image), respectively. Then, $I_{c, \delta}^s(u,v)$ are obtained by convolving the true object-image $I_{c}^o(u,v)$ with $PSF_{c,\delta}(u,v)$ forming the set of misfocused (blurred) color images
\begin{equation}
I_{c,\delta}^{s}(x,y)=PSF_{c,\delta}(x,y)\circledast I_{c}^{o}(x,y),
\label{1}
\end{equation}
where $\circledast$ stays for convolution. In the Fourier domain we have
\begin{equation}
I_{c,\delta}^{s}(f_{x},f_{y})=OTF_{c,\delta}(f_{x},f_{y}) \cdot I_{c}^{o}%
(f_{x},f_{y}).
\label{2}
\end{equation}
The indexes $(o,s)$ stay for object and sensor, respectively.

\begin{figure*}
	\centering
	\includegraphics[width=0.9\linewidth]{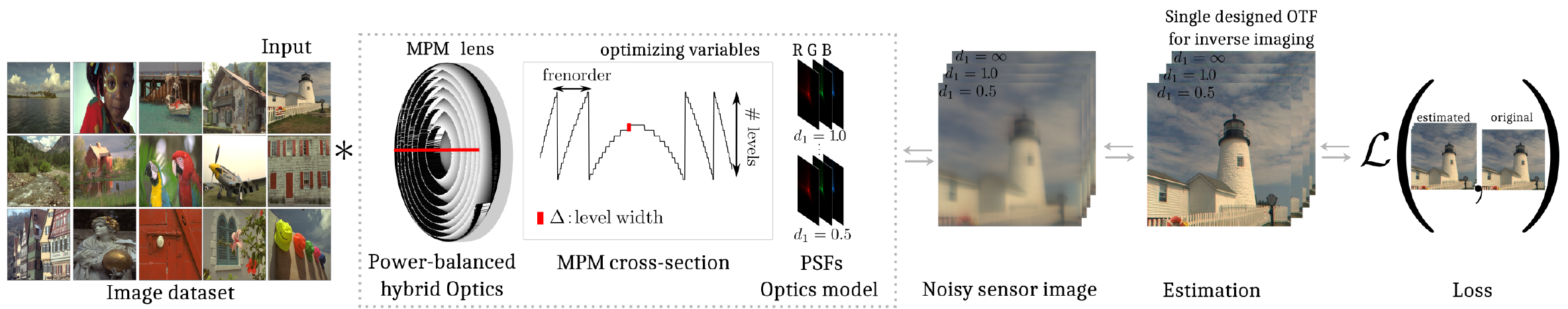}\vspace{-0.5em}
	\caption{\small Differentiable optimization framework of phase-coded optics for achromatic EDoF. The spectral PSFs are convolved with a batch of RGB images. The inverse imaging OTFs provide estimates of the true images. Finally, a differentiable quality loss $\mathcal{L}$, such as mean squared error with respect to the ground-truth image (or PSNR criterion), is defined on the reconstructed images.}\vspace{-1em}
	\label{fig:balancedSystem}
\end{figure*}

\subsubsection{EDoF Image Reconstruction} \label{recon}
For image reconstruction from the blurred data $\{I_{c,\delta}^{s,k}(f_{x},f_{y})\}$, we use a linear filter with the transfer function $H_{c}$ which is the same for any defocus $\delta\in\mathcal{D}$. Let us formulate the design of the inverse imaging transfer function $H_{c}$ as an optimization problem
\begin{align}
\hat{H}_{c} \in \argmin_{H_{c}} \hspace{0.5em}\underbrace{\frac{1}{\sigma^{2}}\sum_{\delta,k,c} \omega_{\delta}||I_{c}^{o,k}-H_{c}\cdot I_{c,\delta}^{s,k}||_{2}^{2}+\frac{1}{\gamma}\sum_{c}||H_{c}||_{2}^{2}}_{J}, 
\label{22_}
\end{align}
where $k\in K$ stays for different images, $I_{c}^{o,k}$ and $I_{c,\delta}^{s,k}$ are sets of the true and observed blurred images (Fourier transformed), $c$ for color, $\sigma^{2}$ stands for the variance of the noise, and $\gamma$ is a Tikhonov regularization parameter. The $\omega_{\delta,0}\leq\omega_{\delta}\leq1$, are the residual weights in the criterion $J$ in \eqref{22_}. We calculate these weights as the exponential function $\omega_{\delta} = exp(-\mu\cdot |\delta|)$ with the parameter $\mu>0$. The norm $||\cdot||_{2}^{2}$ is Euclidean defined in the Fourier domain for complex-valued variables.

Thus, we aimed to find $H_{c}$ such that the estimates $H_{c}\cdot I_{c,\delta}^{s,k}$ would be close to FT of the corresponding true images $I_{c}^{o,k}$. The second summand stays as a regularizer for $H_{c}$. Due to \eqref{2}, minimization on $H_{c}$ is straightforward leading to 
\begin{align}
\hat{H}_{c}(f_{x},f_{y})=\frac{\displaystyle \sum_{\delta \in\mathcal{D}}\omega_{\delta}OTF_{c,\delta}^{\ast}(f_{x},f_{y})}{\displaystyle \sum_{\delta \in\mathcal{D}}\omega_{\delta}|OTF_{c,\delta}(f_{x},f_{y})|^{2}+\frac{reg}{\sum_{k}|I_{c}^{o,k}(f_{x},f_{y})|^{2}}},
\label{solH}
\end{align}
where the regularization parameter $reg$ stays instead of the ratio $\frac{\sigma^{2}}{\gamma}$. The details to derive \eqref{solH} are deferred to Appendix~\ref{app:solH}. In our experiments, we compare inverse imaging for two versions of this Wiener filter. In the first one, the optical transfer function $\hat{H}_{c}$ is defined as above in formula \eqref{solH}. In the second one, we assume, that the sum $\displaystyle \sum_{k}|I_{c}^{o,k}(f_{x},f_{y})|^{2}$ is nearly invariant, and the optical transfer function can be taken the form 
\begin{align}
\hat{H}_{c}(f_{x},f_{y}) =\frac{\displaystyle\sum_{\delta \in\mathcal{D}}\omega_{\delta}OTF_{c,\delta}^{\ast}(f_{x},f_{y})}{\displaystyle \sum_{\delta \in\mathcal{D}}\omega_{\delta}|OTF_{c,\delta}(f_{x},f_{y})|^{2}+reg}.
\label{H_c_reg}
\end{align}
To make a difference between two corresponding inverse imaging procedures, we call the inverse imaging OTF $\hat{H}_{c}$ (Wiener filter) defined by \eqref{solH} and defined by \eqref{H_c_reg} as the inverse imaging OTF (Wiener filter) with invariant and varying regularization, respectively. Sometimes, for simplicity, we call these two procedures as varying or invariant Wiener filters. 

Therefore, the reconstructed images are calculated as 
\begin{equation}
	\hat{I}_{c}^{o,k}(x,y) = \mathcal{F}^{-1}\{ \hat{H_{c}} \cdot I_{c,\delta}^{s,k} \},
\label{misfocus_color_data101}
\end{equation}
where $\mathcal{F}^{-1}$ models the inverse Fourier transform. For the exponential weight $\omega_{\delta} = exp(-\mu\cdot |\delta|)\text{, }\mu>0$ is a parameter that is optimized. The derived OTFs \eqref{solH} and \eqref{H_c_reg} are optimal to make the estimates \eqref{misfocus_color_data101} efficient for all $\delta \in\mathcal{D}$, in this way, we are targeted on EDoF imaging. Note, that ${H_{c}}$ in the form \eqref{H_c_reg} with invariant regularization was proposed in \cite{reza}.

\subsection{MPM Modeling and Design Parameters}
\label{MPM }
In our design of MPM, we follow the methodology proposed in \cite{katkovnik2019lensless}. The details of this design can be seen also in the supplementary material of this paper. The following parameters characterize the free-shape piece-wise invariant MPM: $h$ is a thickness of the varying part of the mask, $N$ is a number of levels, which may be of different height.

\subsubsection{Absolute Phase Model} The proposed absolute phase $\varphi_{\lambda _{0}, \alpha }$ for our MPM takes the form
\begin{align}
\varphi_{\lambda _{0}, \alpha }(x,y) = \frac{-\pi\alpha}{\lambda_0 f_{\lambda_{0}}} (x^{2} + y^{2}) + \beta(x^{3} + y^{3}) +\sum_{r=1,r\not =4}^{R}\rho_{r}P_{r}(x,y),
\label{abs=phase}
\end{align}
where the first term is the quadratic phase, $\alpha$ the optical power-balance between the lens and phase mask for wavelength $\lambda _{0}$, and focal distance $f_{\lambda_{0}}$. The cubic phase of magnitude $\beta$ is a typical component for EDoF, the third group of the items is for parametric approximation of the free-shape MPM using the Zernike polynomials $P_{r}(x,y)$ with coefficients $\rho_{r}$ to be estimated. We exclude the fourth Zernike polynomial (defocus term, i.e., quadratic) because it is considered in the first term. 

The proposed design is a combination of symmetry (quadratic + Zernike polynomials) and non-symmetry (cubic) phase terms. Observe that our parametrization of the optics phase profile significantly differs from works such as \cite{sitzmann2018end} that also considers Zernike polynomials, because the optimizing coefficient $\frac{-\pi\alpha}{\lambda_{0} f_{\lambda_{0}}}$ also affects the contribution in the modulation of the light by the refractive lens employed in our optical system due to $\alpha$. Additionally, since $\alpha$ is the power-balance variable, it deeply affects the physical setup because of the capability to change between hybrid and lensless systems.

\subsubsection{Fresnel Order (thickness)} In radians, the mask thickness is defined as $Q=2\pi m_{Q}$, where $m_{Q}$ is called 'Fresnel order' of the mask which in general is not necessarily integer. Then the phase profile of MPM considering the thickness is calculated as
\begin{equation}
\hat{\varphi}_{\lambda _{0}, \alpha}(x,y) = mod(\varphi_{\lambda _{0}, \alpha }(x,y) + Q/2,Q)-Q/2. 
\label{lens4}
\end{equation}
The operation in \eqref{lens4} returns $\hat{\varphi}_{\lambda _{0}, \alpha}(x,y)$ taking the values in the interval $[-Q/2$, $Q/2)$. The parameter $m_{Q}$ is known as 'Fresnel order' of the mask. For $m_{Q}=1$ this restriction to the interval $[-\pi $, $\pi )$ corresponds to the standard phase wrapping operation.

\subsubsection{Number of Levels} 
The mask is defined on $2D$ grid $(X,Y)$ with the computational sampling period (pixel) $\Delta _{comp}$. We obtain a piece-wise invariant surface for MPM after non-linear transformation of the absolute phase.
The uniform grid discretization of the wrap phase profile $\hat{\varphi}_{\lambda _{0}, \alpha}(x,y)$ to the $N$ levels is performed as
\begin{equation}
	\theta_{\lambda _{0}, \alpha}(x,y) =\lfloor \hat{\varphi}_{\lambda _{0}, \alpha}(x,y) /N \rfloor \cdot N\text{,}
\label{lens5}
\end{equation}
where $\lfloor w \rfloor$ stays for the integer part of $w$. The values of $\theta_{\lambda _{0}, \alpha}(x,y)$ are restricted to the interval $[-Q/2$, $Q/2)$. $Q$ is an upper bound for thickness phase of $\theta_{\lambda _{0}, \alpha}(x,y)$. 

It is worth mentioning that the floor and modulo functions are not differentiable, therefore we use a smoothing approximation to be able of optimizing the thickness and number of levels of MPM. The details of this approximated function can be found in Appendix~\ref{app:smoothing}. The physical size of the mask's pixel is $m_{w}\Delta _{comp}$, where $m_{w}$ is the width of the mask's pixel with respect to the computational pixels. The mask is designed for the wavelength $\lambda_{0}$. Thus, the piece-wise phase profile of MPM is calculated as
\begin{equation}
\varphi_{MPM_{\lambda _{0},\lambda, \alpha}}(x,y) =\frac{\lambda _{0}(n(\lambda )-1)}{\lambda (n(\lambda _{o})-1)}\theta_{\lambda _{0}, \alpha}(x,y),
\label{BPM-final}
\end{equation}
where $\theta_{\lambda _{0}, \alpha}$ is the phase shift of the designed MPM and $n(\lambda)$ is the refractive index of the MPM material, $x\in X, y\in Y$. The MPM thickness $h$ in length units is of the form 
	\begin{equation}
	h_{\lambda _{0}}(x,y)=\frac{\lambda _{0}}{(n(\lambda _{o})-1)} \frac{\theta_{\lambda _{0}, \alpha}} {2\pi }\text{.} 
	\label{lens8}
	\end{equation}
	
\subsection{Optimization Framework}
We develop a framework to optimize the proposed optical system which is summarized in Fig.~\ref{fig:balancedSystem} by stochastic gradient methods with the ADAM optimizer in PyTorch\footnote{The Pytorch library can be downloaded in \url{https://pytorch.org/}. This link also provides proper documentation to correctly use this package.}, an optimized tensor library for Neural Network (NN) learning using GPUs. We express each stage of the model described in the following subsections as differentiable modules.

\subsubsection{Loss Function} Let $\Theta$ be a full set of the optimization parameters defined as
\begin{align}
\Theta= (\alpha,\beta,m_{Q},reg,N, \rho_{r}).
\label{theta}
\end{align}
Then, we use the following multi-objective formulation of our optimization goals
\begin{equation}			
\hat{\Theta}=\argmax_{\Theta}(PSNR(\Theta,\delta), \delta\in\mathcal{D}).
\label{multiobjectivel}
\end{equation}%
In this formulation, we maximize all $PSNR(\Theta,\delta)$, $\delta\in\mathcal{D}$, simultaneously, i.e. to achieve the best accuracy for all focus and defocus situations. Here, $PSNR(\Theta,\delta)$ is calculated as the mean value of $PSNR^k(\Theta,\delta)$ over the set of the test-images, $k\in K$: 
\begin{equation}
	PSNR(\Theta,\delta)=\mean_{k\in K}(PSNR^k(\Theta,\delta)).
	\label{PSNR_total}
\end{equation}
	
There are various formalized scalarization techniques reducing the multi-objective criterion to a simple scalar one. Usually, it is achieved by aggregation of multiple criteria in a single one (e.g. \cite{emmerich2018tutorial}). In this paper, we follow pragmatical heuristics comparing $PSNR(\hat\Theta,\delta)$ as the $1D$ curve functions of $\delta$ to maximize $PSNR(\Theta,\delta)$ for each $\delta\in\mathcal{D}$. Here, $\hat\Theta$ are estimates of the optimization parameter. In this heuristic, we follow the aim of the multi-objective optimization \eqref{multiobjectivel}. The key challenges in developing the proposed optimization framework were to satisfy manufacturing constraints, finding stable optimization algorithms, and fitting models within memory limits.

\subsubsection{Parameters Setup}
We simulate a sensor with a pixel size of $3.45~\mu m$ and a resolution of $512\times 512$ pixels. All Zernike coefficients are initialized to zero at the beginning of the optimization. The optical element is discretized with a $2~\mu m$ feature size on a $3000\times 3000$ grid. In the learning phase, which includes optimizing the optical element and finding the optimal $\alpha,\beta,m_{Q},N$, and $reg$ for the reconstruction, we use a step size of $5\times 10^{-3}$ with Adam stochastic gradient descent solver. Recall that $\alpha$ only can take values over the set $[0,1]$. We limit the study of the number of levels (N) in the interval $[2, 255]$. We range the Fresnel order $m_{Q}$ from 2 to 16. A diameter $D$ of the aperture $P_{A}$ is equal to $6~mm$. A fixed distance between the aperture and the sensor plane $d_{2}$ was chosen to have a sharp image at 1m and distance between the object and the aperture is discretized as $0.5,1.0,2.0,2.4$~m, and $\infty$. The focal distance of the lens is equal to $f=10~mm$, then, the focal distance of the system is $d^0_{1}=1~m$. We experimentally observe that $R=14$ (Zernike coefficients excluding the fourth polynomial) is enough and larger values do not improve image quality significantly. The design wavelength $\lambda_{0} = 510~nm$. After applying the wave-optics image formation model, we also include Gaussian noise in the measurements, with variance equal to $1\times 10^{-4}$. We choose 31 visible wavelengths, in $10~nm$ intervals, in $(400-700)~nm$ to estimate the RGB imaging PSFs. The optimization stage employs 200 epochs, which takes approximately 6 hours on NVIDIA TESLA V 100 with memory of 32GB.

\subsubsection{Dataset} 
We sample images from a dataset of 1244 high-resolution images\footnote{Image databases employed in the training stage can be found in \url{https://data.vision.ee.ethz.ch/cvl/DIV2K/}, and \url{http://cv.snu.ac.kr/research/EDSR/Flickr2K.tar}.} to design the proposed hybrid optics. From this dataset, we present PSNR curves as a function of $d_{1}$ for different Fresnel orders from 2 to 16. Reported PSNRs for each depth $d_{1}$ are calculated as the average over 24 RGB Kodak images \footnote{Kodak dataset can be downloaded in \url{http://r0k.us/graphics/kodak/}.}. 

\subsubsection{Memory constraints} 
Fitting models within memory limits was difficult due to the fine discretization of the optical elements necessary for high fidelity Fresnel propagation. Essential insight for keeping memory usage tractable was to take advantage of the stochastic optimization scheme by randomly binning the depths and wavelengths present in each image into a smaller set. For instance, a given random image is placed over the wide DoF of interest. \vspace{-1em}

\begin{figure*}[t!]
	\centering
	\includegraphics[width=0.7\linewidth]{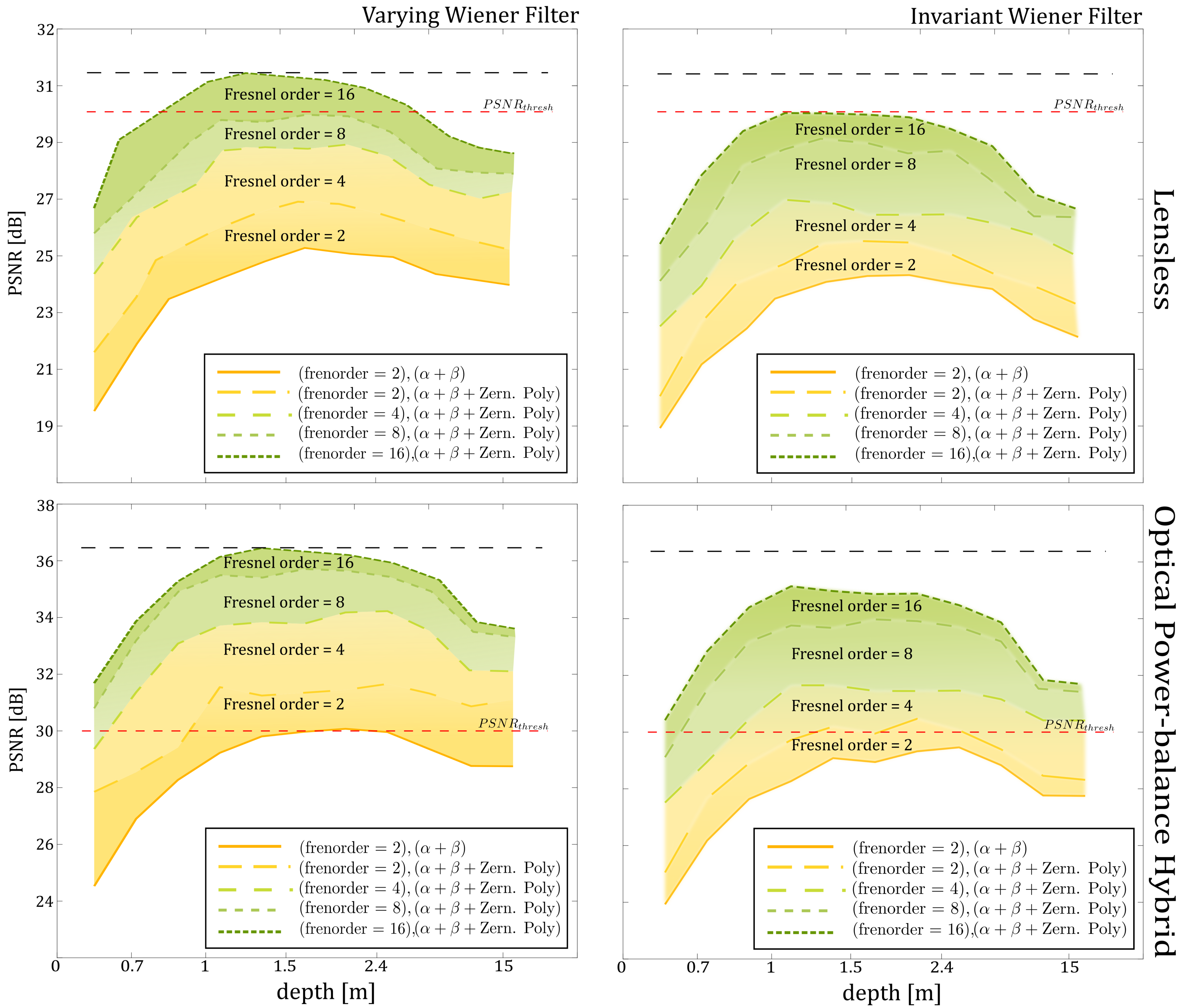}
	\vspace{-0.5em}
	\caption{\small Reconstruction quality (accuracy) in PSNR(dB) versus depth distance $d_{1}$ using the inverse imaging OTFs with invariant/varying regularization for the proposed power-balanced hybrid and lensless systems. The red horizontal lines correspond to the desirable values of $PSNR=30$~dB. The effect of the Fresnel order on reconstruction quality is shown. These numerical results suggest that the highest value of Fresnel order provides the most precise reconstruction. A gain up to 2~dB is obtained by inverse imaging OTF with varying regularization in comparison with the OTF with the invariant regularization. Finally, the power-balanced hybrid overcomes its lensless counterpart up to 5~dB of PSNR.}\vspace{-1em}
	\label{fig:qualitySystemsFresnelOrder}
\end{figure*}

\section{Simulation Tests}\vspace{-0.5em}
\label{simulation-result}
\subsection{MPM parameters effect over power-balanced and lensless systems}
In what follows, the quality of imaging is evaluated by PSNR calculated jointly for RGB channels. Due to the huge amount of possible combinations to analyze the optimizing parameters of MPM, we present the most informative combinations in order to illustrate their importance. For instance, since the thickness of the MPM is an imposed condition by the implementation procedure, we analyze the obtained quality by fixing the Fresnel order and optimizing the rest of the parameters. These results are summarized in Fig.~\ref{fig:qualitySystemsFresnelOrder} for both optical power-balanced hybrid and lensless systems. 

In these figures we present PSNR curves as a function of $d_1$. For the lensless setup, $\alpha$ is equal to 1 according to \eqref{GPF-hybrid}. The best performance is achieved by fixing the Fresnel order and optimizing the remaining parameters for each scenario and optical setup. The colored regions between lines correspond to the obtained quality when the parameters $N$, $\alpha$, and Fresnel order are selected and excluded from the set of parameters subject to optimization, which can be considered as the state-of-the-art strategy. For this second scenario, we choose 1000 random values for $N$ and $\alpha$ (for the lensless case just $N$). These results clearly suggest the advantage of optimizing the number of levels, thickness, and optical power-balance parameters of MPM. 

\begin{figure}[t!]
	\centering
	\includegraphics[width=0.9\linewidth]{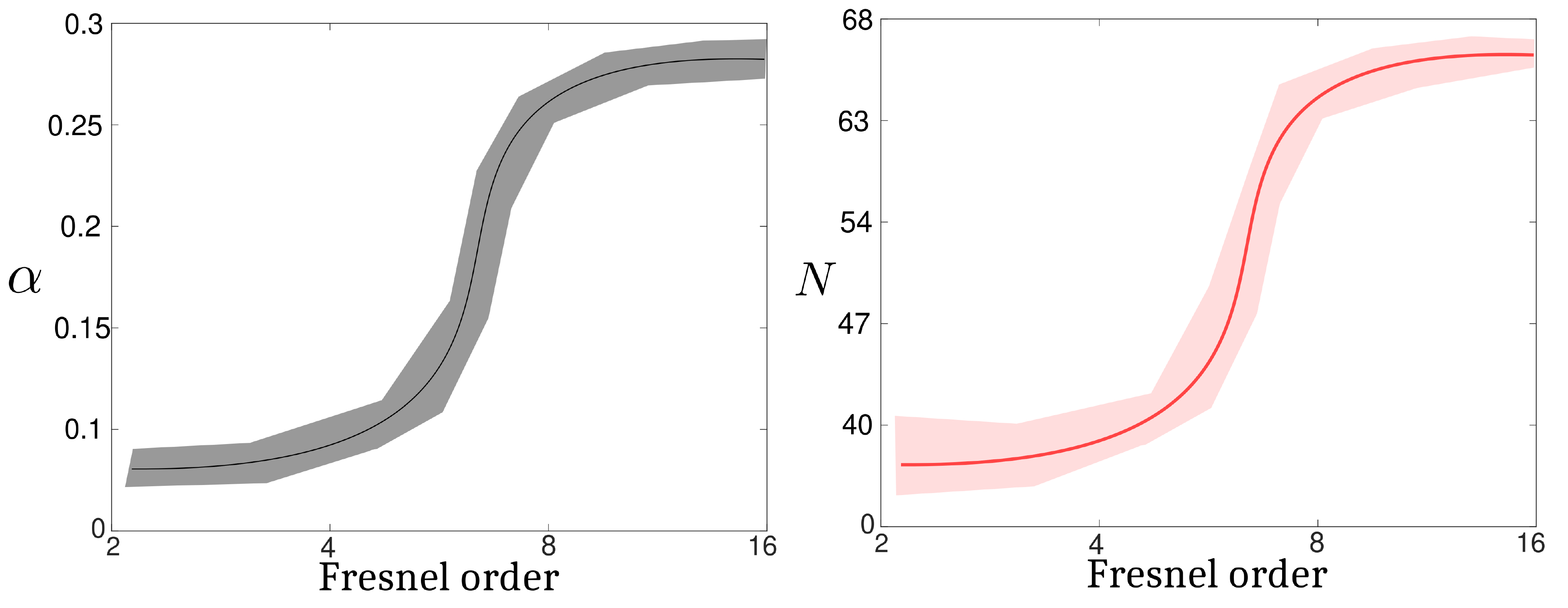}
	\vspace{-0.5em}
	\caption{\small Average of the optimal number of levels (N) and optical power-balance ($\alpha$) values obtained for the proposed system with the end-to-end framework with varying/ invariant regularization versus different Fresnel order values. The solid line stands for the mean along with all the number of levels and power-balance values, and the colored area models the variance. These results suggest a direct dependence between the Fresnel order, the number of levels, and optical power-balance. This dependence reflects that the number of levels and optical power-balance are higher since the Fresnel order increases.}\vspace{-1.4em}
	\label{fig:N_alpha_fresnelorder}
\end{figure}

\begin{figure}[t!]
	\centering
	\includegraphics[width=.9\linewidth]{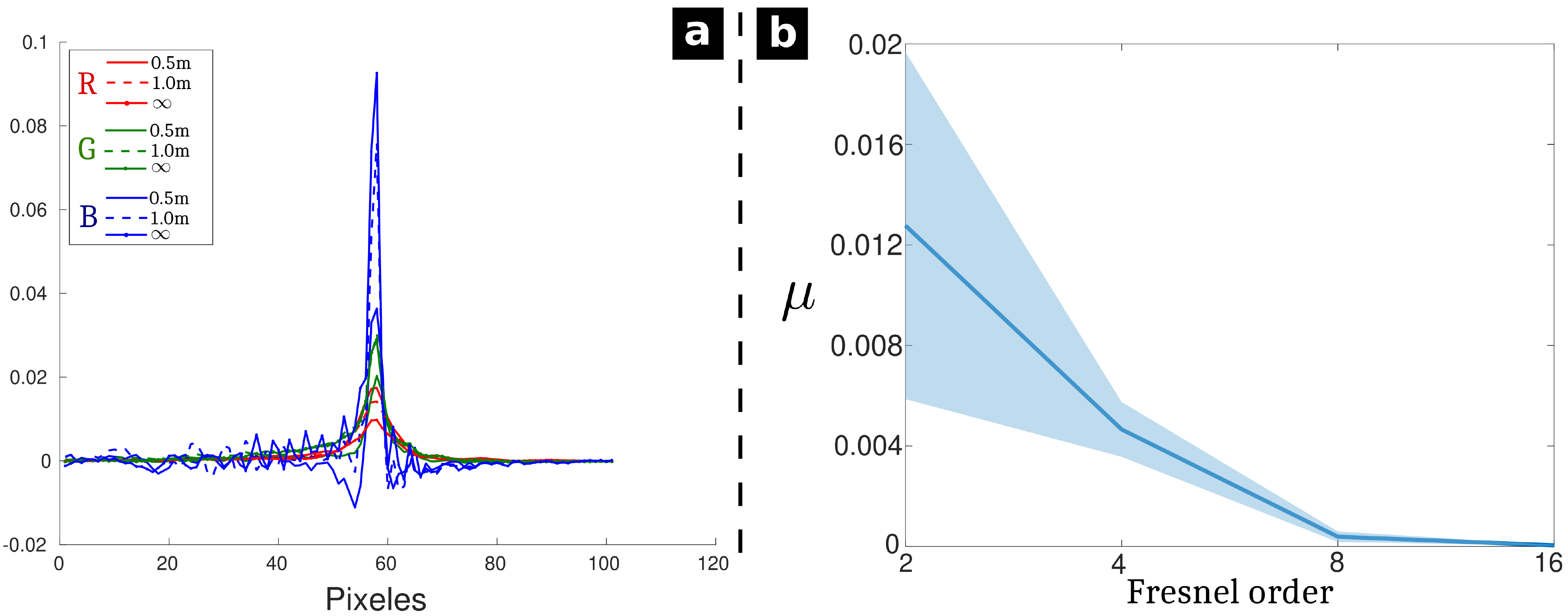}\vspace{-0.8em}
	\caption{\small (a) Central cross-sections of PSF functions of the optical system after deblurring by \eqref{solH}. The ideal deblurring should be close to $\delta$-function, and we observe that the curves are well concentrated around the central point and not far different from each other for all channels and distances. The best result we can see is for the blue color channel. (b) Average of the optimal $\mu$ values obtained for the proposed and lensless systems with the end-to-end framework versus different Fresnel order values. The solid line stands for the mean along all the $\mu$ values and the colored area models the variance. These results suggest a direct dependence between the Fresnel order and the inversion rule in \eqref{misfocus_color_data101} through the value of $\mu$, establishing that $\mu$ can be discarded when the Fresnel order is large enough, meaning that the inversion rule in \eqref{misfocus_color_data101} plays an important role when Fresnel order is small.}\vspace{-1.2em}
	\label{fig:psfs}
\end{figure}

The images in Fig.~\ref{fig:qualitySystemsFresnelOrder} allow multiple comparisons. Note that for each of the presented scenarios, the systems are optimized in end-to-end manner to have a fair comparison of potential of the different optical setups. First, comparing the PSNR curves in the images we may conclude that for the absolute phase with $\alpha$, $\beta$, plus Zernike polynomials, the performance of both optical systems is improving with higher values of PSNRs for larger values of the Fresnel orders varying from 2 to 16. For smaller values of the Fresnel order, this improvement can be very large. For instance, for the optical power-balanced hybrid, we gain about 2-3 dB if the Fresnel order is changed from 2 to 4. At the same time, the change of the Fresnel order from 8 to 16 gives only about 1 dB improvement. It is worth mentioning that the performance gain for Fresnel order larger than 16 is minor and nearly negligible.

\begin{figure*}[t!]
	\centering
	\includegraphics[width=0.75\linewidth]{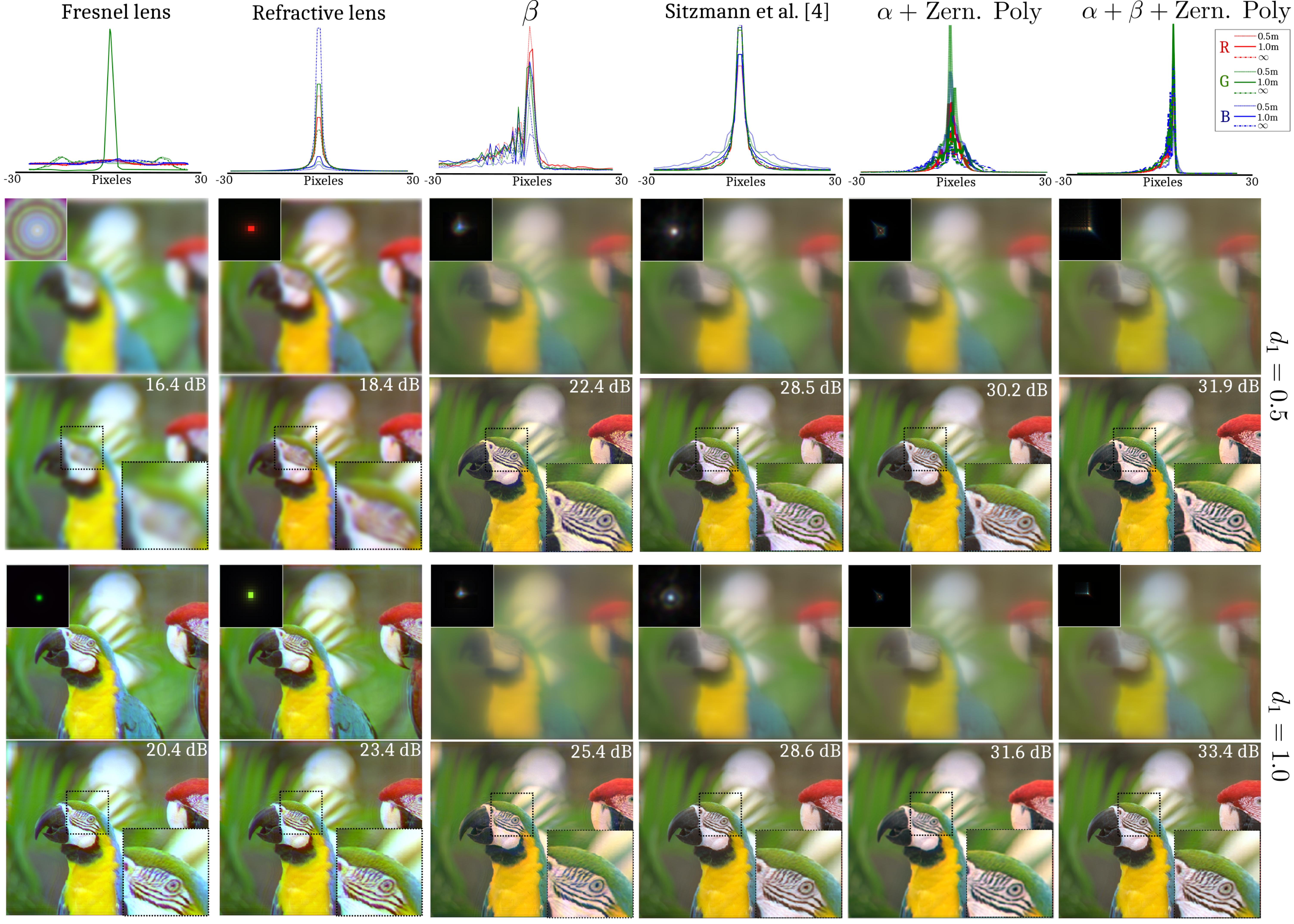}
	\vspace{-0.5em}
	\caption{\small Evaluation of achromatic EDoF imaging in simulation for two depths, $d_{1}=0.5,1.0$, using the proposed mixed OTF in \eqref{solH} in terms of PSNR. We compare the performance of a Fresnel lens optimized for one of the target wavelengths, a refractive lens, lensless setup with cubic MPM (column $\beta$), lensless system with phase profile in \cite{sitzmann2018end}, optical power-balanced hybrid without cubic component (column $\alpha + \text{Zern. Poly}$) and the proposed system (column $\alpha + \beta + \text{Zern. Poly}$) with Fresnel order equal to 16 and $\alpha=0.22$ following results in Fig.~\ref{fig:qualitySystemsFresnelOrder}. The resulting PSFs are shown for all target wavelengths and depths. Sensor images and PSFs (rows 2 and 4) exhibit strong chromatic aberrations for the Fresnel lens, refractive lens, and the lensless system with cubic MPM, whereas the lensless system in \cite{sitzmann2018end}, power-balanced hybrid without cubic component and the proposed optics mitigate wavelength-dependent image variations much better where our optics has the best performance (the curves are better concentrated around the focal point). After deconvolution (rows 3 and 5), the sharpest images are obtained with the proposed optics.}\vspace{-1em}
	\label{fig:simulated}
\end{figure*}

Second, comparing the left and right columns in Fig.~\ref{fig:qualitySystemsFresnelOrder} we may note that the OTFs with the Wiener varying regularization demonstrate the improved performance of about 1 dB as compared with the counterpart with the invariant regularization for both optical setups. In the second row, for the optical power-balanced hybrid, the peak of the upper curve for OTF with the varying regularization case is about 1 dB more than the corresponding PSNR value for OTF with the invariant regularization. Third, the PSNR curves in each image are given for different absolute phase models for the design of MPM, marked by the parameters $\alpha, \beta,$ and Zernike polynomials. Comparing these curves, we may evaluate the advantage of more complex models. In particular, the optimized Zernike polynomials demonstrate visible improvements in performance for all algorithms. Fourth, the comparison of the lensless system versus the proposed optical power-balanced hybrid is definitely in favor of the latter one with the great advantage of about 5 dB.
	
Overall, the absolutely best results are clearly demonstrated by the proposed system with phase profile defined by including $\alpha, \beta,$ and Zernike polynomials \eqref{abs=phase}, OTFs with Wiener varying regularization \eqref{solH}, and Fresnel order equal to 16. Regarding the achieved DoF, we note that for the proposed hybrid with power-balanced optical power, DoF covers the design interval $[0.5,\infty]~m$ with the clear advantage of the reconstruction with varying regularization. The lensless system fails to achieve a similar result. It can be also shown that the lens-only system is still far from reaching this goal. 

We finalize the exposition of the numerical results related to the optimized MPM parameters by presenting the behavior of the designing framework for the number of levels and optical power-balance versus Fresnel order. These results are summarized in Fig.~\ref{fig:N_alpha_fresnelorder}. We can observe a clear dependence between the Fresnel order and these two variables suggesting the need of the number of levels ($N$) and the optical power-balance ($\alpha$) in the optimization pipeline, indicating in this case that $N$ and $\alpha$ need to be higher when the Fresnel order increases for the particular task of extended-depth-of-field. These experiments reveal that lensless or just lens systems (without optical power-balance) are sub-optimal optical solutions for EDoF since the obtained range for $\alpha$. \vspace{-0.6em}

\subsection{Inverse Imaging}
To explain the results shown in Fig.~\ref{fig:qualitySystemsFresnelOrder}, we analyze the behavior of PSF functions and OTF designed for inverse imaging. Fig.~\ref{fig:psfs}(a) provides additional confirmation that indeed the effect of color imaging without chromatic aberrations is successfully achieved. In Fig.~\ref{fig:psfs}(a), the cross-sections of PSF functions of the proposed system after deblurring by \eqref{solH} are shown for three RGB color channels and three different distances. This deblurring process is calculated as $\mathcal{F}^{-1}\{ \hat{H_{c}} \cdot OTF_{c,\delta}\}$, for $\delta=0.5,~1.0,$ and$~\infty$. The curves are shown for the proposed system with Fresnel order equal to 16 and $\alpha=0.22$ following Fig.~\ref{fig:N_alpha_fresnelorder}. The curves are well concentrated around the focal point and consolidated with respect to each other suggesting high-quality imaging for all color channels.

Here, we also study the relation between $\mu$ and the Fresnel order obtained under the end-to-end optimization. Therefore, we want to determine how the Fresnel order value affects the performance of the mixed OTF through the variable $\mu$. Recall that $\mu$ is needed to estimate the scene when it is composed of several objects simultaneously located at different distances and defines the weights $\omega_{\delta}$ for the designed OTF. The desired analysis between Fresnel order and $\mu$ is summarized in Fig.~\ref{fig:psfs}(b). Specifically, this figure presents the average of the optimal $\mu$ values obtained for the proposed and lensless systems with the end-to-end framework. These optimal values were estimated in the experiments performed for Fig.~\ref{fig:qualitySystemsFresnelOrder}. The solid line, in Fig.~\ref{fig:psfs}(b), stands for the mean along all the $\mu$ values and the colored blue area models the variance. These results suggest a direct dependence between the Fresnel order an the inversion rule in \eqref{misfocus_color_data101} through the value of $\mu$. This dependence reflects that $\mu$ can be discarded when the Fresnel order is large enough. That is, the inversion rule in \eqref{misfocus_color_data101} play an important role when Fresnel order is small (less than 8).\vspace{-0.5em}


\subsection{Simulated Reconstructions}
We present simulated reconstructions for two depths $d_{1}=0.5$ and $1.0~m$ using the proposed mixed OTF in \eqref{solH} summarized in Fig.~\ref{fig:simulated}. This figure analyzes the Fresnel lens optimized for one of the target wavelengths, a refractive lens, lensless setup with cubic MPM (column $\beta$), lensless system with phase profile in \cite{sitzmann2018end}, optical power-balanced hybrid without cubic component (column $\alpha + \text{Zern. Poly}$), and the proposed system (column $\alpha + \beta + \text{Zern. Poly}$) with Fresnel order equal to 16, $\alpha=0.22$, and the number of levels equal to 62 following results in Fig.~\ref{fig:N_alpha_fresnelorder}. From these results, we observe that the resulting point spread functions are shown for all target wavelengths and depths. Sensor images and PSFs (rows 2 and 4) exhibit strong chromatic aberrations for the Fresnel lens, refractive lens, and the lensless system with cubic MPM, whereas the lensless system in \cite{sitzmann2018end}, power-balanced hybrid without cubic component and the proposed optics mitigate wavelength-dependent image variations much better where our optics has the best performance (the PSF curves are better concentrated around the focal point). After deconvolution (rows 3 and 5), the sharpest images are obtained with the proposed optics. Additionally, combining the results from columns, we clearly observe the motivation of using the phase profile in \eqref{abs=phase}. 

\begin{figure}[t!]
	\centering
	\includegraphics[width=0.7\linewidth]{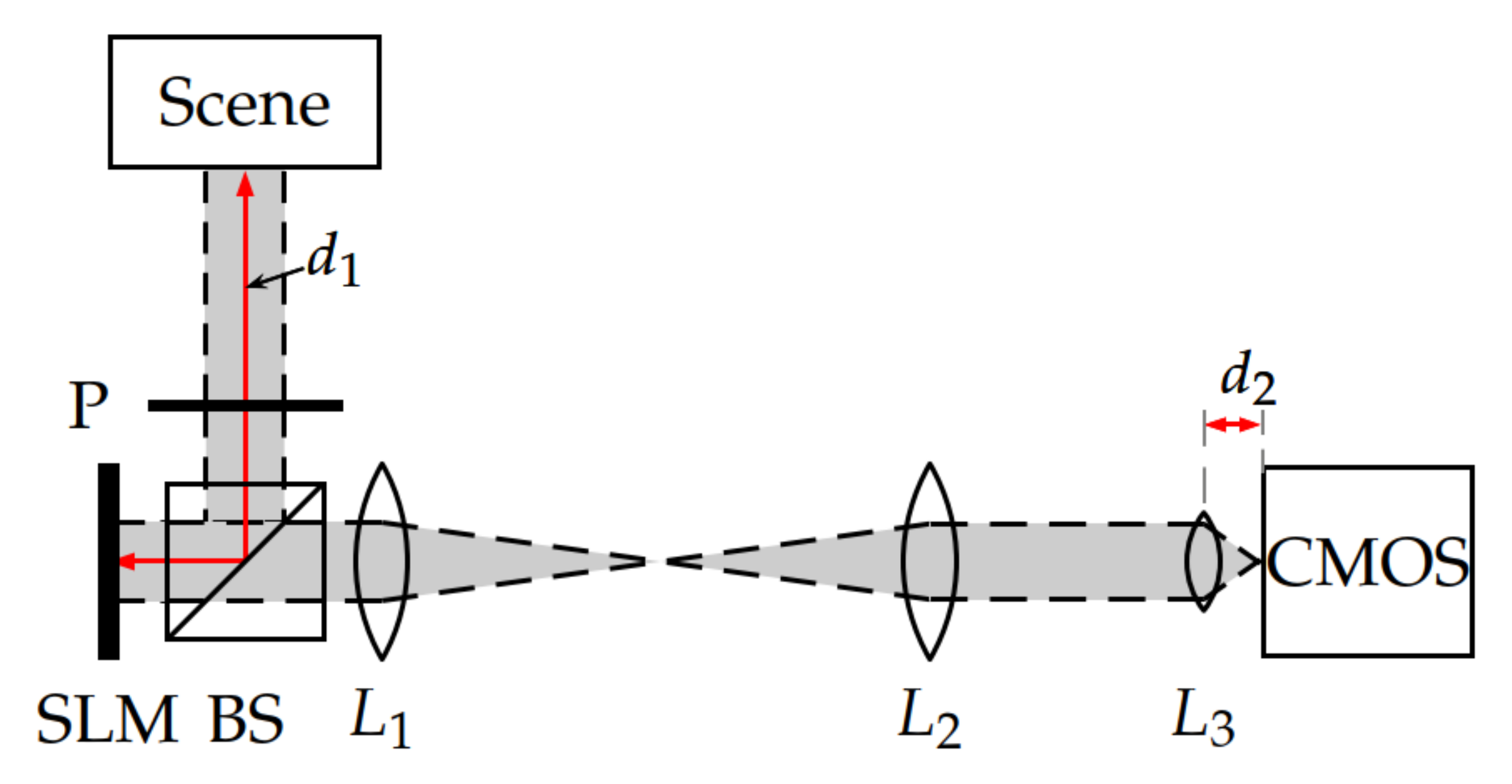}
	\caption{\small Experimental setup. P is a polarizer, BS is a beamsplitter, SLM is a spatial light modulator. The lenses $L_1$ and $L_2$ form the 4f-telescopic system projecting wavefront from the SLM plane to the imaging lens $L_3$, CMOS is a registering camera. $d_1$ is a distance between the scene and the plane of the hybrid imaging system 'Lens~$ \&$~MPM', and ~$d_2$ is a distance between this optical system and the sensor. }\vspace{-1.6em}
	\label{fig:setup}
\end{figure}

\section{Experimental Tests}
\label{experimental-result}
\subsection{Optical Setup and Equipment}
We point out that due to the new included variables in the MPM design (number of levels, thickness, and power-balance), the number of possible combinations to compare the proposed system with the state-of-the-art is huge and in order to avoid building several MPM to physically analyze the performance of our camera, we build a setup which exploits the phase capabilities of a spatial light modulator (SLM) to investigate the performance of the proposed MPM design. The optical setup is depicted in Fig.~\ref{fig:setup}, where 'Scene' denotes objects under investigation; the polarizer, 'P', keeps the light polarization needed for a proper wavefront modulation by SLM; the beamsplitter, 'BS', governs SLM illumination and further light passing; the lenses '$L_1$' and '$L_2$' form a 4f-telescopic system transferring the light wavefront modified by SLM to the lens '$L_3$' plane; the lens '$L_3$' forms an image of the 'scene' on the imaging detector, 'CMOS'. 

For MPM implementation, we use the Holoeye phase-only GAEA-2-vis SLM panel, resolution $4160\times2464$, pixel size $3.74~\mu$m, '$L_1$' and '$L_2$' achromatic doublet lenses with diameter $12.7$~mm and focal distance of $50$~mm, BK7 glass lens '$L_3$' with diameter $6$~mm and focal distance $10.0$~mm; 'CMOS' Blackfly S board Level camera with the color pixel matrix Sony IMX264, $3.45~\mu$m pixels and $2448\times2048$ pixels. This SLM allows us to experimentally study the optical power-balanced hybrid with the phase distribution of the designed MPM (implemented on SLM) additive to the imaging lens ‘$L_3$’. The MPM phase was created as an 8-bit \textit{*.bmp} file and imaged on SLM. We calibrated the SLM phase delay response to the maximum value of $3.6\pi$ for a wavelength of $510$~nm. This $3.6\pi$ corresponds to the value 255 of \textit{*.bmp} file for the phase image of MPM. 

\begin{enumerate}
	\item \textbf{Imaging: }Test images in 'scene' plane are displayed on a screen with $1440\times2960$ pixels and $570$ppi. The distance $d_{1}$, between the screen and SLM is varied from $0.5$m to $2$m. The distance $1.0$m is the focal point and the others for the defocus. This distance range is enough to study and experimentally prove the advantages of the proposed system in terms of sharpness and low chromatic aberrations. \vspace{-0.5em}
		
	\item \textbf{SLM and MPM design: } We recall that the SLM only provides a phase delay of up to $3.6\pi$ which is a limiting aspect to fully study the effect of the designing variable Fresnel order. In fact, the simulation tests show that larger values of the Fresnel order result in better imaging. Despite this limitation, this work studied two cases, when Fresnel order=1.2 and 1.4. Nevertheless, even with these small values, we demonstrate high-quality imaging and advantage of the higher value of Fresnel order.\vspace{-0.5em}
		
	\item \textbf{PSF acquisition: } To calibrate the system in Fig.~\ref{fig:setup}, we use a fiber of diameter $200\mu m$ as a point-source for white light in a dark room. Additionally, the PSFs are downsampled to resolution of $3.74 \mu m$ of resolution to fit the pitch size of the employed SLM. From these acquired PSFs, we produce experiments for the ten distances $d_{1}$ equal to $0.5,~0.65,~0.8,~0.9,~1.0,~1.15,~1.3,~1.5,~1.8$, and 2.0m, and use these estimates to compute the deblurring invariant/varying OTF following the formulas \eqref{solH} and \eqref{H_c_reg}.\vspace{-0.3em}
		
	\item \textbf{Reconstruction algorithm:} To estimate the scene, the experimental PSFs and blurred images are needed as input. Then, the deblurring OTF either invariant or varying formula is computed following \eqref{H_c_reg} and \eqref{solH} to estimate the image using \eqref{misfocus_color_data101}. After this step, a denoising process equipped with a sharpening procedure \cite{Dabov07jointimage} is performed over the estimated scene to improve the quality of imaging. This final denoised image is returned as the estimated scene from experimental data.\vspace{-0.5em}
	\end{enumerate}

\subsection{Experimental Results}
We present observations and images reconstructed from the observed blurred measurements for depth range 0.5-2.0 meters two Fresnel orders ($1.2$ and $1.4$) using the varying/invariant Wiener filtering methods as in \eqref{solH} and \eqref{H_c_reg}, following the optical setup described in Fig.~\ref{fig:setup}. We present the experimental and simulated PSFs for each system and different distances employing the process acquisition for the physical setup described in the previous subsection. This experimental data is acquired for the proposed optimized optical power-balanced hybrid, lens+cubic phase MPM, lensless with cubic phase MPM, and lens systems with the interest of validating the simulated results provided in Section \ref{simulation-result}. The optimal value for the power-balance variable $\alpha$ is $0.05$. The estimated images are obtained using the reconstruction algorithm summarized described above. The OTF step in this algorithm demands the choice of the $reg$ value following \eqref{solH} and \eqref{H_c_reg} that we select by cross-validation to obtain the best visual quality for the reconstructed image in the interval $[10^{-5},10^{-3}]$.

\begin{figure*}[t!]
	\centering
	\includegraphics[width=0.8\linewidth]{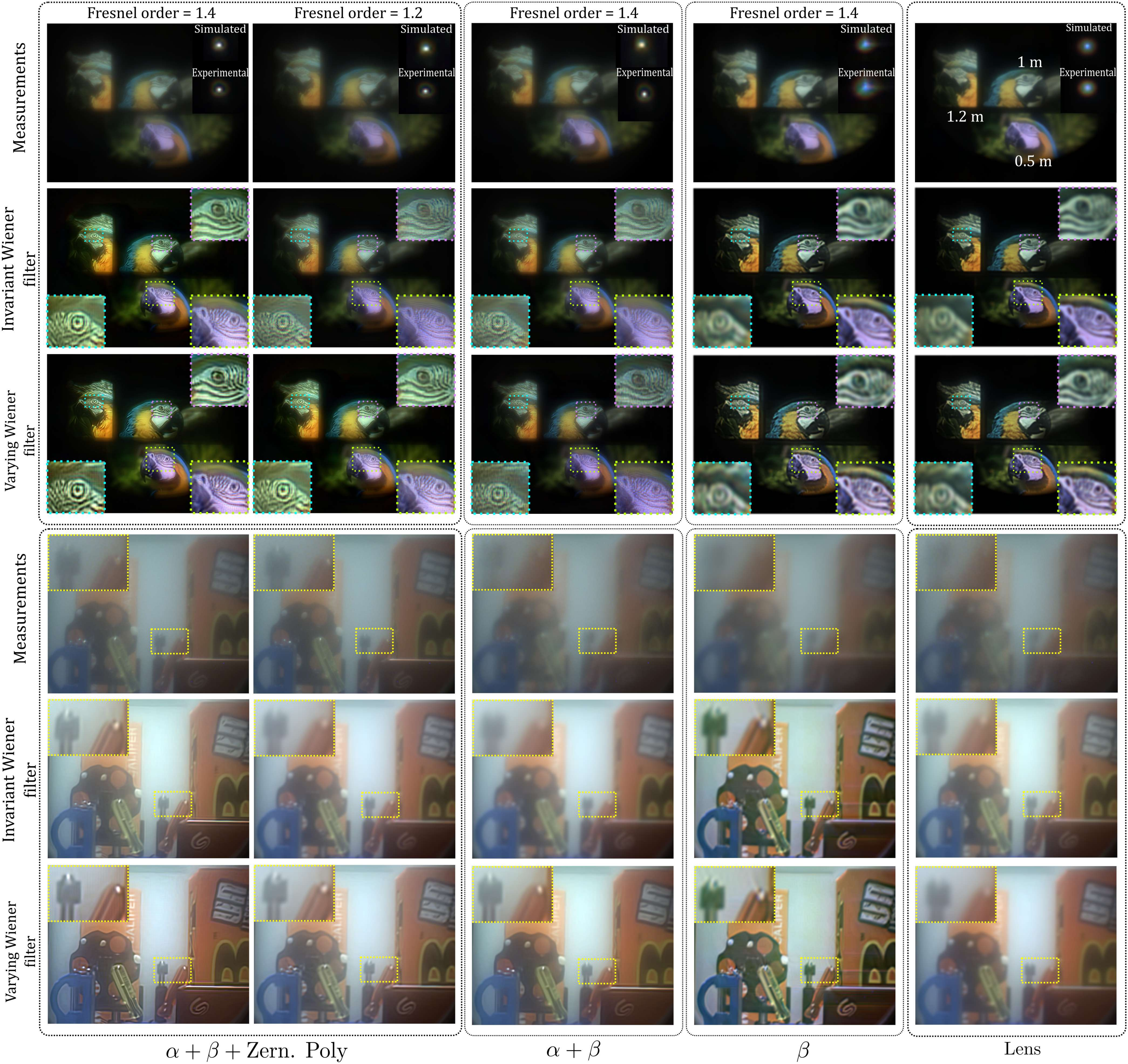} \vspace{-0.8em}
	\caption{\small Reconstructed images from experimental blurred measurements that contain three objects at three different distances $d_{1}=0.5,1.0$, and $1.2$ m, a casual real scene (depth range 0.5-2.0m), and two Fresnel orders are compared for four optical setups: proposed power-balanced hybrid (column $\alpha + \beta +$ Zern. Poly), lens + MPM corresponding to cubic companion (column $\alpha + \beta$), lensless with cubic phase MPM (column $\beta$), and lens only (column Lens). To estimate the scene, we use varying/invariant Wiener filtering as in \eqref{solH} and \eqref{H_c_reg} for distances $d_{1}=0.5,0.65,0.8,0.9,1.0,1.15,1.3,1.5,1.8$, and 2.0m. The experimental and simulated PSFs are shown in the measurement rows. The imaging results confirm that the highest value of Fresnel order provides the best reconstruction quality and that the advantage of the varying Wiener filtering in \eqref{solH} over its invariant version in \eqref{H_c_reg} is a mitigation of the chromatic aberrations and reduction of the noise. Additionally, we verify that the optimized proposed system is superior to its competitors in terms of sharpness and chromatic aberrations suggesting the effectiveness of the optimization framework described in Fig.~\ref{fig:balancedSystem}. Lastly, we confirm the ability of the optimized OTF in \eqref{solH} to estimate the longitudinal scene.}\vspace{-1em}
	\label{fig:realData3}
\end{figure*} 
	
In this experiment, the first scene contains three objects (three parrot's images) located simultaneously at three different distances $d_{1}=0.5,1.0,$and $1.2 m$, and the second is a casual real scene  (depth range 0.5-2.0m). We acquire the mixed blurred measurements to determine the effectiveness of the optimized OTF in \eqref{solH}. These results are summarized in Fig.~\ref{fig:realData3}. Besides, to complement these experiments, we provide supplementary material that contains the behavior of the proposed system for different values of $\alpha$. In this case, to estimate the scene, we use varying/invariant Wiener filtering with the optimized OTF in \eqref{solH} and \eqref{H_c_reg} showing also the experimental and simulated PSFs for all the systems where the value of $\mu$ in the formula of the weights $\omega_{\delta}$ in \eqref{22_} is fixed as $1\times10^{-3}$. Notice that the effect of $\mu$ in the reconstruction quality can be only analyzed when the scene is composed of several objects simultaneously located at different distances as is the case for Fig.~\ref{fig:realData3}. In this scenario, the reference images (column Fresnel order$=1.4$, rows 'varying Wiener filter') suggest that the highest value of Fresnel order provides the best reconstruction quality which is obtained for Fresnel order = 1.4 compared with Fresnel order = 1.2, in terms of sharpness and low chromatic aberrations. 

The inserted zoomed images in Fig.~\ref{fig:realData3} reveals that the lens+cubic MPM (column $\alpha + \beta$), lensless with cubic MPM (column $\beta$), and lens systems have strong chromatic aberrations. In fact, it is now clear that the advantage of the varying Wiener filtering in \eqref{solH} over its invariant version in \eqref{H_c_reg} is the mitigation of the chromatic aberrations and reduction of the noise since the proposed OTF partially corrects these issues for the lens+cubic MPM, lensless with cubic MPM, and lens systems and in parallel suggesting the superiority of the proposed optical power-balanced hybrid system. Therefore, we have that the returned design optics from the optimization framework described in Fig.\ref{fig:balancedSystem} indeed provides achromatic EDoF behavior. Lastly, we verify the effectiveness of the optimized OTF in \eqref{solH} to estimate the longitudinal scene, and the smoothing function to model the number of levels and Fresnel order of MPM. More results can be found in the Supplemental information. Finally, we consider that a future research direction of \eqref{solH}, \eqref{H_c_reg} can be dynamic imaging because of their computational efficiency and scalability.\vspace{-0.8em}
	
\section{Conclusion}
\label{sec:conclusion}
It is shown in this paper that the optimized power-balanced hybrid optical system composed from refractive lens and diffractive phase coding MPM in the scenario of achromatic EDoF imaging demonstrates advanced performance as compared with the two counterparts: the single refractive lens and the lensless system with MPM as an optical element. The optical power-balance in the proposed hybrid and the design of MPM both optimized in the end-to-end framework are crucial elements of this advance. The algorithm for multi-objective optimization balances PSNR's values for imaging with different defocus distances and in this way enables EDoF imaging. The designed hybrid optics is insensitive to defocus and in this way automatically enables achromatic imaging as these PSFs are insensitive also to the dispersion of spectral characteristics of MPM and the lens. One of the original elements of this paper is the OTF (in two version with invariant and varying regularization) optimal for inverse imaging in EDoF scenarios. We show also that the Fresnel order (thickness) and number of levels of MPM are of important design parameters suggesting the effectiveness of the smoothing function. To the best of our knowledge, it is an original observation. At least, we have not seen this sort of statement concerning design of DOEs. The advanced performance of the proposed optical setup is demonstrated by numerical simulation and experimental tests. For our implementation of MPM, we use a high-resolution spatial light modulator (SLM). As further work, we consider design and implementation of the optical power-balanced hybrid camera with a thick MPM.

\begin{backmatter}
\bmsection{Acknowledgments} This work is supported by the CIWIL project funded by Jane and Aatos Erkko Foundation, Finland. 
\bmsection{Disclosures} The authors declare no conflicts of interest.
\bmsection{Supplemental document}
See Supplement 1 for supporting content. 
\end{backmatter}

\appendix
\section{Solution of $\eqref{22_}$}
\label{app:solH}
Observe that the criterion $J$ can be rewritten as
\begin{equation}
J=\frac{1}{\sigma^{2}}\sum_{\delta,k,c} \omega_{\delta}||I_{c}^{o,k}-H_{c}\cdot OTF_{c,\delta}\cdot I_{c}^{o,k}||_{2}^{2}+\frac{1}{\gamma}\sum_{c}||H_{c}||_{2}^{2}.\label{222}%
\end{equation}

The minimum condition for $J$ is calculated as $\frac{\partial J(f_{x},f_{y})}{\partial H_{c}^{\ast}(f_{x},f_{y})}=0,$ where (*) stays for complex-conjugate. After the derivative calculation and some manipulations, we arrive at the equation:
\begin{align}
&\frac{1}{\sigma^{2}}\sum_{\delta} \omega_{\delta} H_{c}(f_{x},f_{y}) \cdot|OTF_{c,\delta}(f_{x},f_{y})|^{2}\cdot \sum_{k}|I_{c}^{o,k}(f_{x},f_{y})|^{2} \nonumber\\
-&\frac{1}{\sigma^{2}}\sum_{\delta} \omega_{\delta}OTF_{c,\delta}^{\ast}(f_{x},f_{y})\cdot \sum_{k}|I_{c}^{o,k}(f_{x},f_{y})|^{2} +\frac{1}{\gamma}H_{c}(f_{x},f_{y}) =0 
\label{eq1}
\end{align}
with the solution for $H_{c}$
\begin{align}
\hat{H}_{c}(f_{x},f_{y})=\frac{ \displaystyle \sum_{\delta \in\mathcal{D}}\omega_{\delta}OTF_{c,\delta}^{\ast}(f_{x},f_{y})}{ \displaystyle \sum_{\delta \in\mathcal{D}}\omega_{\delta}|OTF_{c,\delta}(f_{x},f_{y})|^{2}+\frac{\sigma^{2}}{\gamma \sum_{k}|I_{c}^{o,k}(f_{x},f_{y})|^{2}}}
\label{H_c}
\end{align}
Here $\displaystyle \sum_{k}|I_{c}^{o,k}(f_{x},f_{y})|^{2}/\sigma^{2}$ is a signal-to-noise ratio typical for the Wiener filter. It is assumed that $\displaystyle \sum_{k}|I_{c}^{o,k}(f_{x},f_{y})|^{2}>0$. We use this solution in the form:
\begin{align}
\hat{H}_{c}(f_{x},f_{y})=\frac{\displaystyle \sum_{\delta \in\mathcal{D}}\omega_{\delta}OTF_{c,\delta}^{\ast}(f_{x},f_{y})}{\displaystyle \sum_{\delta \in\mathcal{D}}\omega_{\delta}|OTF_{c,\delta}(f_{x},f_{y})|^{2}+\frac{reg}{\sum_{k}|I_{c}^{o,k}(f_{x},f_{y})|^{2}}},\label{H_c2} 
\end{align}
where the regularization parameter $reg$ stays instead of the ratio $\frac{\sigma^{2}}{\gamma}$. This $reg$ is used for tuning the filter replacing $\gamma$.

\begin{figure}[t!]
	\centering
	\includegraphics[width=0.75\linewidth]{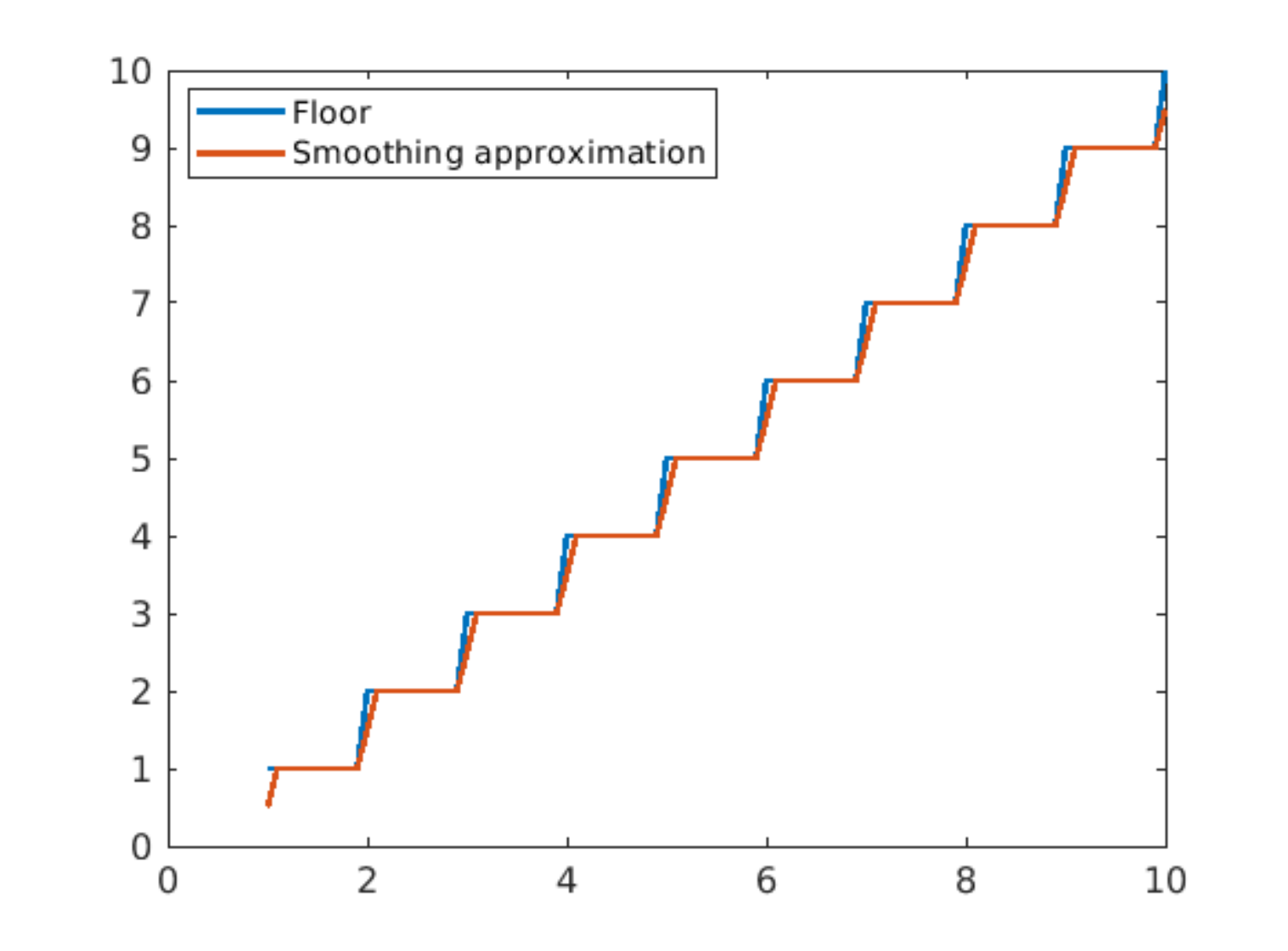}\vspace{-1em}
	\caption{\small Comparison between the smoothing approximation of the floor function described in \eqref{eq:floor}. From this figure we observe that \eqref{eq:floor} provides a close estimation of the true $\lfloor w \rfloor$ value.}\vspace{-1.5em}
	\label{fig:floor}
\end{figure}
\section{Smoothing Approximation Functions}
\label{app:smoothing}
In order to have a fully differentiable end-to-end design framework optimizing the number of levels and Fresnel order of MPM, we take advantage of the following two smoothing functions to approximate the floor and round operations. Specifically, we approximate $\lfloor w \rfloor$ in \eqref{lens5} as
\begin{equation}
	\lfloor w \rfloor \approx g(w) = w - \frac{1}{2} - \frac{1}{\pi}\tan^{-1}\left(\frac{-0.9999\sin(2\pi w)}{1-0.9999\cos(2\pi w)}\right).
	\label{eq:floor}
\end{equation}
The quality of the above approximation is summarized in Fig.~\ref{fig:floor}. From this figure we observe that \eqref{eq:floor} provides a close estimation of the true $\lfloor w \rfloor$ value.

Now, to approximate the round function we use 
\begin{equation}
	\text{round}(w) \approx g(w+0.5).
	\label{eq:round}
\end{equation}
We remark that \eqref{eq:round} is used to approximate the non-differentiable operation in \eqref{lens4}. Specifically, we have 
\begin{equation}
	mod(w,z) \approx w - zg(w/z).
\end{equation}
The quality of $g(w)$ to approximate $\text{round}(w)$ is also summarized in Fig.~\ref{fig:floor} (since it depends on \eqref{eq:floor}).\vspace{-0.5em}

\bibliography{sample}

\bibliographyfullrefs{sample}



\end{document}